\documentclass[%
 reprint,
 amsmath,amssymb,
 aps,
]{revtex4-1}

\usepackage{graphicx}
\usepackage{dcolumn}
\usepackage{bm}
\usepackage{epstopdf}
\usepackage{xcolor}

\newcommand{\ndyvo}{Nd$^{3+}$:YVO$_4$}

\newcommand{\dhdhh}{$^4\!I_{9/2}-\,^4F_{3/2}\,$}

\begin{document}

\title{Photon echoes in optically dense media}

\author{Sergey A. Moiseev$^{1}$}
\author{Mahmood Sabooni$^{2,3}$}
\author{Ravil V. Urmancheev$^1$}%
\affiliation{%
$^1$Kazan Quantum Center, Kazan National Research Technical University n.a. A.N.Tupolev-KAI, 10 K. Marx, Kazan 420111, Russia}
\affiliation{$^2$Institute for Quantum Computing, Department of Physics and Astronomy, University of Waterloo, Waterloo, Ontario, N2L 3G1, Canada}
\affiliation{$^3$Department of Physics, University of Tehran, 14399-55961, Tehran, Iran}

\date{\today}

\begin{abstract}

Coherent nonlinear multi-pulse processes, nonlinear waves and echo effects in resonant media are the topical problems of modern optics and important tools of coherent spectroscopy and quantum information science.
We generalize the McCall-Hahn area theorem to the formation of an arbitrary photon echo generated during the multi-pulse excitation of the optically dense resonant media.
The derived theorem made it possible to reveal the nonlinear mechanism of generation and evolution of the photon echo signals inside the media after a two-pulse excitation.
We find that a series of self-reviving echo signals with total area of $2\pi$ or $0\pi$ is excited and propagates in the media depth, with each pulse having an individual area less than $\pi$.
The resulting echo pulse train is a new alternative to the well-known soliton or breather.
The developed pulse-area approach paves the way for more precise coherent spectroscopy, studies of different photon echo signals  and quantum control of light pulses in the optically dense media.

\end{abstract}

\maketitle

Studies of coherent multi-pulse nonlinear effects like photon echo and four-wave mixing open wide opportunities for understating of light-atom interactions, fundamental processes of nonlinear and quantum optics, provide powerful techniques for spectroscopic investigation of atoms and molecules and are considered as a principal tool for implementation of basic processes in practical quantum information science \cite{Yetzbacher2007,Christensson2008,Dorfman2016,Pezz2018,Mourou2019}.
Herein, photon echo technique \cite{KopvillemPhEcho1963,Kurnit1964} attracts an especial everlasting attention in coherent spectroscopy \cite{Kurnit1964} and light pulse storage 
\cite{HEER197749,Samartsev1980,1981JEPTStyrkov,Mossberg:82,Carlson:831}.
Recently, the  photon echo in optically dense media opened promising  opportunities for quantum storage of a large number of light pulses 
\cite{MoiseevKroll2001,Tittel2009,Lvovsky2009,SparkesNatComm2011,Usmani2010} and quantum  processing \cite{OpticalQMapplicationReview}  that determined a steady interest and  elaboration of numerous protocols of photon echo based quantum memory \cite{Tittel2009,Hosseini2011,Rani2017,Minnegaliev_2018,Saglamyurek2018,Guo2019,Mazelanik2019}, which are 
important for the creation of quantum repeater \cite{RevModPhys.83.33}, microwave quantum memory \cite{PhysRevLett.105.140503,Moiseev2018}, etc.

The study of the properties of a two- and three-pulse  photon echoes in optically dense media is  the main task in the development of the multi-pulse spectroscopy and photon echo quantum memory schemes in such media.
The most general theoretical description of the coherent resonant interaction of multi-pulse light fields with resonant atoms can be provided by the pulse area theorem \cite{McCallHahn1969,Lamb1971,allen1975optical,Eberly:98,PhysRevLett.88.243604,Chaneliere:14,PhysRevA.92.063815,ThreePulseAreaTheorem}.
In early works on the two-pulse (primary) photon echo, it was found that the initial excitation could result in the generation of multiple echo signals 
\cite{HAHN1971265,allen1975optical} followed by a long-term investigation of the underlying mechanism \cite{FRIEDBERG1971285,Lamb1971,allen1975optical,Moiseev1987,1998-Azadeh-PRA,1998-Wang-OC,1999-Wang-PRA,PhysRevA.79.053851,Li2010,Tsang:03}.
Quite early an analytic solution for total area of all the echoes was obtained \cite{HAHN1971265,allen1975optical,1998-Azadeh-PRA}, that proved that the total pulse area can tend asymptotically towards $2\pi$ in the media depth if the initial pulse area of two exciting laser pulses exceeds $\pi$.
However, this solution does not allow one to describe the behavior of each individual echo pulse.

Previously acquired solution for the primary echo pulse area predicted that the echo pulse area never exceeds $\pi$ and generally decays in the depth of the media \cite{Moiseev1987} .
This finding again stressed the ambiguity of the known physical picture behind the formation of the total nonlinear response to the multi-pulse excitation.
In the recent years the stakes were raised by the demand for an efficient optical solid-state quantum memory and the noted interest in coherent multi-pulse interactions in the optically dense media.

In this Rapid Communication we find an analytical solution of the photon echo pulse area theorem posed in \cite{HAHN1971265,FRIEDBERG1971285,Lamb1971} in 1971.
By analysing the solution we for the first time discover the mechanism of self-induced transparency \cite{McCallHahn1969} for two- and many-pulse excitation of the atomic media leading to the formation of many echo pulses.
To do that we find the general analytic solution for the pulse area of an arbitrary secondary photon echo signal.
The found solutions show that the echo signals are excited coherently one after another in a certain area of the medium and then disappear, generating new echo signals and creating a self-reviving echo sequence.
We show that depending on the input pulse areas this echo pulse train forms a multi-pulse analogue to the well-known single pulse $2\pi$ optical soliton or a $0\pi$ optical breather despite each individual echo pulse area never exceeding $\pi$.
Herein, by using the highly non-linear nature of the light-atom interaction we can control the total response of the media.  
Being near the thresh-old, when the incoming area of the second pulse is close to $\pi$,  and  by  slightly  changing  it  to  being $<\pi$ or $> \pi$ one can initiate a huge change in the outcome from an optical soliton to an optical breather, respectively. 
This also demonstrates the potential of the pulse area approach for coherent spectroscopy of the optically dense media. 

First we reproduce the McCall-Hahn area theorem and derive the general equation for the pulse area of an arbitrary  echo pulse starting with the usual reduced set of Maxwell-Bloch equations \cite{allen1975optical}  for the light field and atomic system:
	\begin{align}
	\begin{split}
	[ \partial_z + c^{-1}\partial_t ] \Omega & = i \frac{\mu}{2} \langle P\rangle, 
	\\
    \partial_t u & = - \Delta v - \gamma u,
	\\
	\partial_t v & = \Delta u - \gamma v + \Omega w,
	\\ 
	\partial_t w & = -\Omega v,\\
	\end{split}
	\label{eq:mb_set}
	\end{align}
    where $\vec{r} = \vec{r}(t,z,\Delta) = (u,v,w)^T$ is the Bloch vector, each component depending on time $t$, spatial coordinate $z$ and atomic detuning $\Delta$;  $P = u-iv$ - atomic polarization, electric field $E(t,z) =  \varepsilon(t,z) \exp[i(kz-\omega t)] + c.c.$ is described by a complex light field envelope $\varepsilon(t,z)$ with corresponding Rabi frequency $\Omega (t,z) = (2 d/\hbar) \varepsilon (t,z)$; $\mu = 4\pi N d^2\omega/\hbar c$, $\gamma = 1/T_2$, $T_2$ is the coherence lifetime of the atomic transition and $\langle...\rangle\equiv \int_{-\infty}^{\infty}  G(\Delta) ...d\Delta$ is the averaging over the inhomogeneous broadening. From now on for simplicity, we do not denote the existing dependence on $z$ in atomic and field variables $\vec{r}$ and $\Omega$.
    
	By transferring to the pulse area $\theta = \int_{-\infty}^\infty dt \Omega (t)$  and follow \cite{McCallHahn1969,Eberly:98} to find that incoming pulse areas $\theta_1, \theta_2$ satisfy the well-known pulse area theorem:
	\begin{equation}
	\partial_z \theta = \tfrac{1}{2}\alpha w_0(z)\, \sin \theta(z), 
	\label{eq:area1}
	\end{equation}
	where $w_0$ is the initial inversion of the atomic system, $\alpha$ is the resonant absorption coefficient \cite{allen1975optical}. 
	The first pulse propagates in the undisturbed media, with $w_0=-1$ and partially inverts for the second pulse, so $w_0= -\cos\theta_1$.
	Substituting $w_0$ into Eq.~\eqref{eq:area1} we get the well-known solutions \cite{allen1975optical}:
	\begin{align}
	\begin{split}
	   	\theta_1(z) & = 2 \arctan \left[e^{-\alpha z/2} \tan \dfrac{\theta_1(0)}{2}\right],
	    \\
	    \theta_2(z) & = 2\arctan \left[ \kappa ~\mathrm{sech} \left( \beta -  \frac{\alpha}{2}z \right)\right],
	\end{split}
    \label{eq:th_two_solution}
	\end{align}
    where $\beta=\ln\{\tan[\frac{\theta_1(0)}{2}]\}$ and $\kappa=\tan[\frac{\theta_2(0)}{2}]/\sin[\theta_1(0)]$. 

    Eqs. \eqref{eq:area1} and \eqref{eq:th_two_solution}  can be used to find the total area of all excited photon echoes \cite{HAHN1971265,FRIEDBERG1971285,allen1975optical,1998-Azadeh-PRA}: 
    \begin{equation}
        \theta_{\Sigma e}(z) = 2 \arctan \left[e^{-\alpha z/2} \tan \tfrac{\theta_{1}(0)+\theta_{2}(0)}{2}\right]-\theta_2(z)-\theta_1(z).
       \label{eq:sum_area}
    \end{equation}

This solution predicts that if $\theta_2(0)<\pi, \theta_1(0)+\theta_2(0) > \pi$, the total area of all echo pulses asymptotically tends to $2\pi$ \cite{HAHN1971265}.
It leaves however a lot of uncertainty about the mechanism and physics of the photon echo generation, since any information about the particular photon echo signals remains hidden. 
How exactly different echoes combine into $2\pi$ pulse area? 
What is the contribution of an individual echo? 
Moreover, if input pulse areas $\theta_1(0)<\pi/2, \theta_2(0) >\pi$, Eq.~\eqref{eq:sum_area} predicts the sum of all echoes to be $0$. 
What happens with the different echo signals in this case and does that mean that there will be no echoes? 
To answer all these questions, we have to analyze the generation of each echo signal individually. 
        
To find the area theorem for an arbitrary individual photon echo signal we integrate the first of Eqs.~\eqref{eq:mb_set} over time around the time of echo emission $t_e$, from $t_0 = t_e-\tau/2$ to $t_1=t_e+\tau/2$, where $\tau$ is the delay between the pulses.
We should also clarify the time scales assumed for the following derivation. 
Firstly, we assume non-overlapping pulses $\tau \gg \delta t_{1,2}$ with pulse duration being much smaller than coherence time $\delta t_i \ll T_2$, $i=1,2,e1,...$, to neglect the relaxation during the pulses.
Secondly, inhomogeneous broadening of the atomic system is much larger that the pulse spectrum $\Delta_{in} > 1/\delta t_{1,2}$.
Thirdly, for simplicity we consider a solid state system, meaning $T_1\gg T_2$ and thus we can neglect the population decay between the pulses. 
In short, $1/\Delta_{in} < \delta t_{1,2} \ll \tau \lesssim T_2$.

The expressions under the integrals, $P_0(z,\Delta)$ and $w(t,z,\Delta)$ are complex expressions consisting of several oscillating components.
However most of these components will give $0$ after averaging over $\Delta$ in Eq. \eqref{eq:mb_set}.
To find the proper expression for the echo area we need to only take into account the  phasing components of polarization and inversion that contribute to the echo formation.
The details of the integration and equation handling can be found in the Supplemental material.

As a result we obtain the general equation for an arbitrary echo pulse area:
\begin{equation}
\partial_z \theta (z)= \frac{1}{2}\alpha [ 2 v_0(z) \cos^2\frac{\theta(z)}{2} + w_0(z)\sin\theta(z)],
\label{eq:area_general}
\end{equation}
where $w_0(z)$ and $v_0(z)$ are the initial values ($t=t_e-\tau/2$) of the Bloch vector resonance components with $\Delta = 0$ which only give nonzero response in the field equation in (1). 
After transition to $\eta = \tan\tfrac{\theta(z)}{2}$ we get  a linear equation $\partial_z \eta(z) = \frac{\alpha}{2}[v_0(z)+w_0(z) \eta(z) ]$ with clear solution.

Equation ~\eqref{eq:area_general}  describes the pulse area of a chosen echo signal given the phasing coherence  $v_0$ in a presence of spectral uniform inversion $w_0$ and  
Eq. ~\eqref{eq:area_general} comes down to finding $v_0(z)$ and $w_0(z)$ for each echo signal.
In Supplemental material we describe the algorithm that allows to find the $v_0, w_0$ for an arbitrary echo. 
But whatever they may be, we note that  $|\theta|$ never exceeds $\pi$.
Below we investigate the analytic solutions for the pulse areas of all the echo signals. 

For primary echo we have $\vec{r}(t)=U(t-\tau)T(\theta_2)U(\tau)T(\theta_1)\vec{r}(0)$, $t_0 = 3\tau/2$ and the correct phasing components of $\tilde{v}_{0}(3\tau/2), \tilde{w}_{0}(3\tau/2)$ \cite{Moiseev1987,2019-OptExpress}:
    \begin{align}
    \begin{split}
        v_{0}(3\tau/2,z) & = \Gamma_{\tau}^2 \sin\theta_1(z)\sin^2\tfrac{\theta_2(z)}{2},
        \\
        \tilde{w}_{0}(3\tau/2,z) & = -\cos\theta_1(z)\cos\theta_2(z),
    \end{split}
    \label{eq:primary_vw}
    \end{align}
    where $\Gamma_{\tau} = e^{-\gamma\tau}$ is the relaxation term.
    Corresponding Eq.~\eqref{eq:area_general} gives primary photon echo pulse area:
    \begin{equation}
        \theta_{e1}(z) =2\arctan\left[\Gamma_{\tau}^2 \sin\theta_1(0) \sin^2 {\tfrac{\theta_2(z)}{2}}\sinh\tfrac{\alpha z}{2} \right].
        \label{eq:echo_area_solution}
    \end{equation}

After the incoming pulses and the primary echo pulse we have $\vec{r}(t)=U(t-2\tau)T(\theta_{e1})U(\tau)T(\theta_2)U(\tau)T(\theta_1) \vec{r}(0),~ t_0=5\tau/2$ and the phasing components $v_{0}(5\tau/2,z), w_{0}(5\tau/2,z)$ are:
\begin{align}
\begin{split}
v_{0} =  v_{01} & + v_{02} = 
\tfrac{1}{2} \Gamma_\tau^2 \sin\theta_1(z) \sin\theta_{e1}(z) \sin\theta_2(z)
\\
     + & \Gamma_\tau^2 \cos\theta_1(z)\sin^2\tfrac{\theta_{e1}(z)}{2} \sin\theta_2(z),
        \\
w_{0}  = w_{01} & + w_{02} =  -\Gamma_\tau^2 \sin\theta_1(z)\sin^2\tfrac{\theta_2(z)}{2} \sin\theta_{e1}(z)  
\\
 - & \cos\theta_1(z)  \cos\theta_2(z)\cos\theta_{e1}(z).
\end{split}
\label{eq:2nd_echo_v_w}
\end{align}
    
The first terms in both equations $v_{01} (z)= \tfrac{1}{2}\Gamma_\tau^2 \sin\theta_1\sin\theta_2\sin\theta_{e1}$ and $w_{01}(z) = - \Gamma^2_\tau \sin\theta_1\sin^2\tfrac{\theta_2}{2}\sin\theta_{e1}$ are proportional to $\sin\theta_1(z)$ and vanish when the first pulse is absorbed. 
They are responsible for stimulated photon echo generated by incoming pulses and the primary echo pulse. 
The other two components $v_{02}(z) = \Gamma_\tau^2\cos\theta_1\sin\theta_2\sin^2\tfrac{\theta_{e1}}{2}$ and $w_{02}(z) = -\cos\theta_1\cos\theta_2\cos\theta_{e1}$ are proportional to $\cos\theta_1$ are correspond to the secondary two-pulse photon echo created by the second pulse and the primary echo pulse. 

Analysis of the successive echoes  follows the same procedure but requires more calculations since $v_{0}$ and $w_{0}$ have more terms with each step.
In the Supplemental material we introduce the phasing polarization and inversion components for the third and the fourth echoes and discuss the physical meaning of different contributions.
It is obvious that the described procedure can be applied for the case with comparable transverse and longitudinal relaxations and for other light-atom equations.   

\begin{figure}
        \centering
        \includegraphics[width = .8\linewidth]{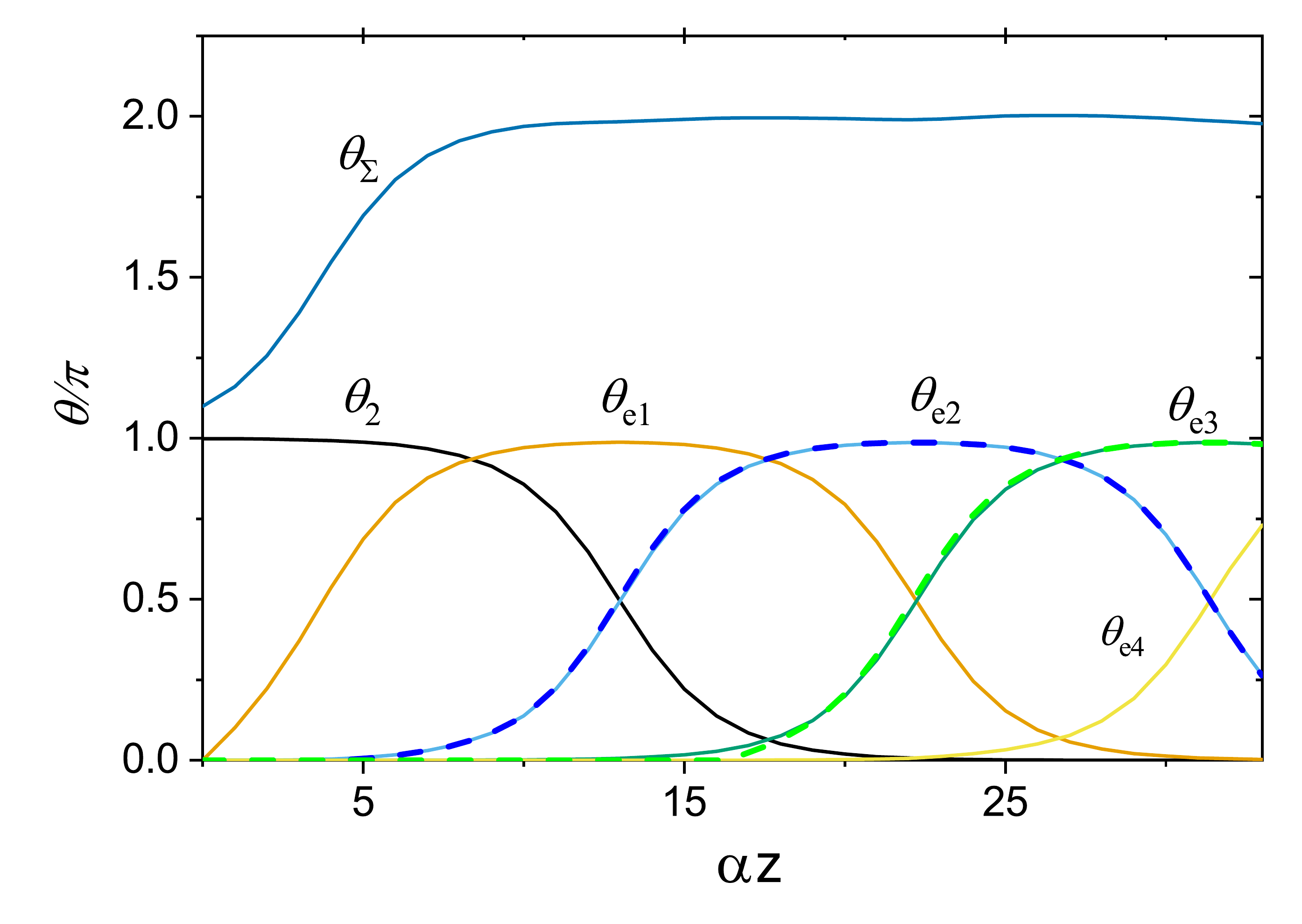}
        \caption{The multi-pulse excitation in an optically dense medium. Incoming pulse areas are $\theta_1(0) = 0.1\pi, \theta_2(0) = 0.999\pi$. The dashed lines show the approximate solution for the second echo $\theta_{e2}$ ($\alpha z_1 = 4.1$, blue dashed line) and the third echo $\theta_{e3}$($\alpha z_2=16.3$, green dashed line).}
        \label{fig:echo_areas_999}
    \end{figure}

We will now proceed to clarify the mechanism of the total $2\pi$ pulse area formation when $\theta_1(0)+\theta_2(0)>\pi$. 
Figure \ref{fig:echo_areas_999} shows the spatial behavior of the area of incoming pulses, echo pulses and the total area depending on the optical density of the medium for $\theta_1(0)=0.1\pi, \theta_2(0)= 0.999\pi$. 
We see that incoming pulses excite primary and secondary echoes that in turn excite subsequent echos. 
Each echo pulse is born, propagates and eventually dies out within a finite spatial interval.
However the total area of all existing pulses behaves strictly in accordance with McCall-Hahn area theorem  Eq.~\eqref{eq:sum_area} and remains close to $2\pi$.
This is realized due to the precise spatial consistency of all the echoes involved.

\begin{figure}
\centering
\includegraphics[width = .8\linewidth]{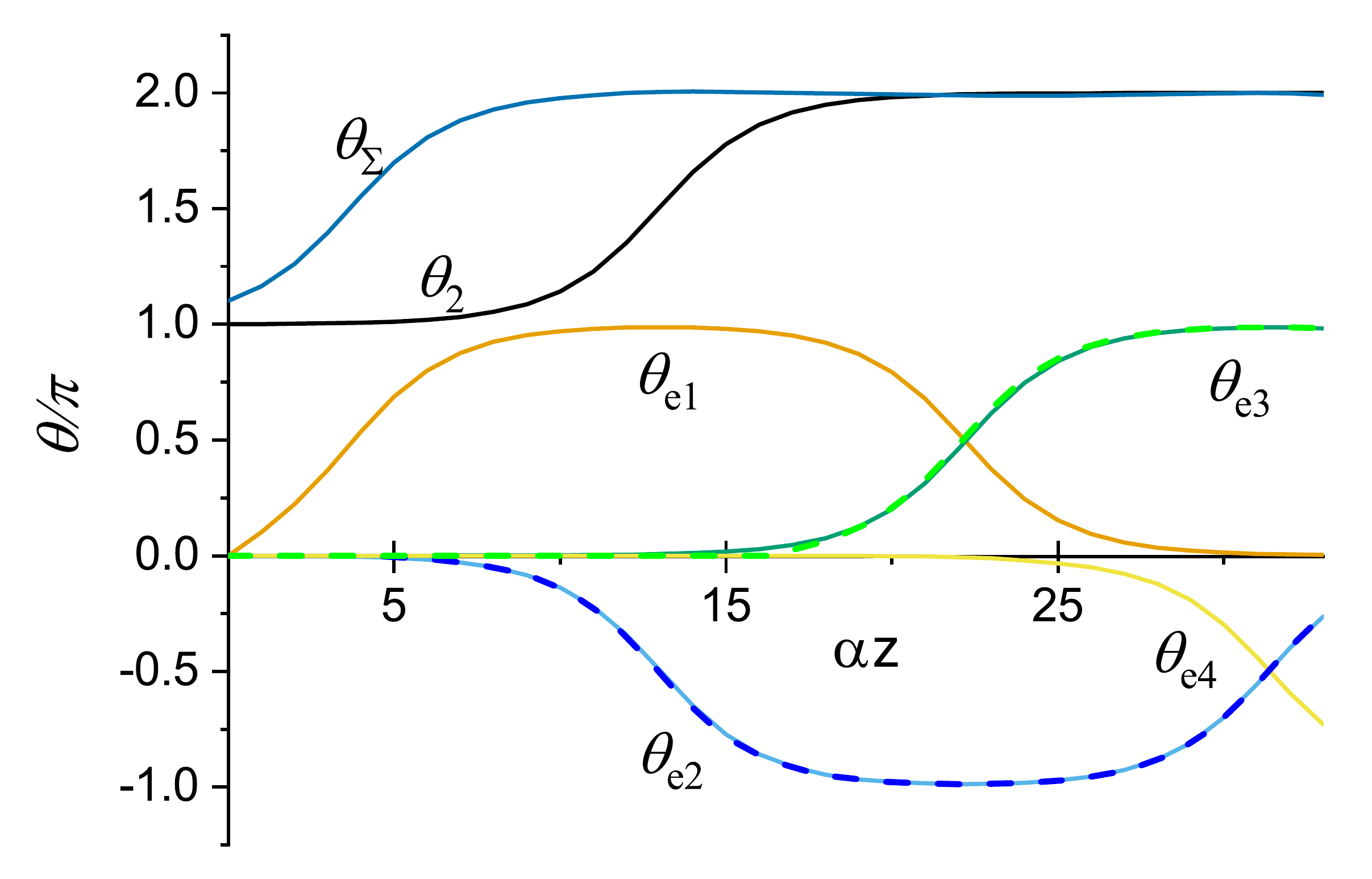}
\caption{Evolution of the multi-pulse excitation in an optically dense medium. Incoming pulse areas are $\theta_1(0) = 0.1\pi, \theta_2(0) = 1.001\pi$. The dashed lines show the approximate solution for the second echo $\theta_{e2}$ ($\alpha z_1 = 4.1$, blue dashed line) and the third echo $\theta_{e3}$($\alpha z_2=16.3$, green dashed line).}
\label{fig:echo_areas_1001}
\end{figure}

The case of $\theta_2(0)>\pi$ really helps to highlight the benefits of looking at an individual echo signal rather than at the sum of all echo signals. 
The second incoming pulse is big enough to form a $2\pi$-soliton on its own, and McCall-Hahn area theorem predicts that the sum of all echoes will equal $0\pi$. 
The impression could be that after some point in the medium there are no echoes at all.
The real picture however is much more vivid, there are many hidden echoes with nontrivial areas working together to comply with the McCall-Hahn area theorem.
Figure~\ref{fig:echo_areas_1001} showcases this echo pulses' behavior for $\theta_1(0)=0.1\pi, \theta_2(0)=1.001\pi$.
Each two of the subsequent echoes have opposite phases, so they are canceling each other in a dynamical equilibrium, resulting in $0\pi$ total pulse area at any point of the medium. 
Figure \ref{fig:echo_areas_1001} also shows that the primary echo assists the formation of the $2\pi$ total area, which would otherwise happen much further into the medium.


We note that the echo areas in Figs.~\ref{fig:echo_areas_999},\ref{fig:echo_areas_1001} behave very similar, differing only in their spatial delays.
This is the case, when we can neglect the stimulated echo terms in Eqs.~\eqref{eq:2nd_echo_v_w} and find a highly accurate approximate analytic solution for each pulse area.
For example, we write for the secondary echo area ($z>z_1$):
\begin{equation}
\tan \frac{\theta_{e2}}{2} =\Gamma_{\tau} \sin\theta_2(z_1) \sin^2 {\frac{\theta_{e1}(z)}{2}}\sinh\tfrac{\alpha}{2} (z-z_1), 
\label{eq:approx_sol}
\end{equation}
where $\theta_{e1}$ is given in Eq.~\eqref{eq:echo_area_solution} with the initial pulse areas taken at the transition point $z_1$:  $(\theta_1(0),\theta_2(0)) \rightarrow (\theta_2(z_1),\theta_{e1}(z_1))$.
By doing so we assume that at $z=z_1$ the first pulse was successfully absorbed by the media and  neglect polarization and inversion components acquired at $z<z_1$. 
solution for $\theta_{e2}$ is shown with dashed lines in Figs.~\ref{fig:echo_areas_999},\ref{fig:echo_areas_1001}.

Equations \eqref{eq:echo_area_solution} and \eqref{eq:approx_sol} describe the pulse area at the output of the optically dense media. 
Moreover, given $\delta t_1 > \delta t_2$ they can also accurately describe the peak energy of the echo pulse \cite{1999-Wang-PRA,2019-OptExpress}.
This easy to measure quantity can be used for coherent multi-pulse spectroscopy of the optically dense media, where usual spectroscopy is complicated due to strong nonlinear light-atoms interaction.
In this highly nonlinear regime the conventional Beer law $I_{echo} = I_0 \Gamma_\tau^2$ is not valid while Eq. \eqref{eq:echo_area_solution} can be used to measure $\Gamma_\tau$ dependence.

It also is interesting to discuss the experimental detection of photon echo train generation and what it can lead to.
As it is seen in Figs.~\ref{fig:echo_areas_999},\ref{fig:echo_areas_1001}, one can  experimentally observe only $2$ or $3$ light pulses at the output of the optical density medium, while other pulses will be highly suppressed.
Herein in media with higher optical densities, we will see  only higher order echo pulses, characterized experimentally by later arrival times. 
The photon echo experiments in such media are quite typical for many quantum memory protocols.
In particular, interesting opportunity is to try detecting the spatial evolution of photon echo inside such media, for example in the rare-earth ions doped crystals \cite{Tittel2009,CHANELIERE201877,Hua_2018}.

One possible candidate for high optical density and large Rabi frequency is \dhdhh transition of \ndyvo$\:$  at $897.705\: \text{nm}$ with dipole moment $d = 9.16\times 10^{-32}\: \text{C.m}$.
Considering $P=100\: mW$ and beam radius of $r=1\mu$m one could reach up to $\Omega \sim 250\: \text{MHz}$.
The $\pi$-pulses can be as brief as several nanoseconds which is much shorter than $T_2$.
These pulses are spatially squeezed in the medium up to 4 orders of magnitude by the group velocity reduction in the presence of a spectral hole in the optical transition \cite{Sabooni2013b}, this would allow to observe spatial evolution of the solitons and echo pulses inside the medium.

It is worth noting   that only soliton-like pulses can propagate through the medium without changing their temporal form and transferring atoms to their initial state.
Accordingly, the photon echo pulses in the generated train will be stretch in time and ultimately overlap with each other deep in the medium  forming a single $2\pi$-soliton in case of Fig.~\ref{fig:echo_areas_999}.
Similarly the stretching echo pulses will asymptotically form a $0\pi$-breather, for the case of Fig.~\ref{fig:echo_areas_1001}. 
In the core of these transformations lies conservation laws of Maxwell-Bloch equations \cite{Lamb1972}.


Finally, we summarize and conclude the long-lasting derivation of the two-pulse photon echo area theorem started over 45 years ago in \cite{HAHN1971265,FRIEDBERG1971285,Lamb1971}, providing an analytic solution for the pulse area of any desired photon echo signal.
We showcase the power of the pulse area approach by exploring the rich physics behind the two-pulse echo excitation of an optically dense medium in two previously understudied cases: $\theta_1(0) < \pi, \theta_2(0) \lesssim(\gtrsim) \, \pi$.
For the first time we demonstrate that in both these cases a self-reviving echo train is excited deep in the medium with total pulse area $2\pi$ in the first case and $0\pi$ in the second previously unknown case.
Thus a slight change in the second pulse area can lead to the dramatic change in the nonlinear multi-pulse media response: an optical soliton in one case or a soliton followed by a breather in the other case. 
At the same time the complex spatial dynamic of the total nonlinear media response after the two-pulse excitation is precisely aligned with the general McCall-Hahn area theorem prediction.

The developed approach of photon echo pulse area theorem can provide new insights in general analysis of coherent multi-pulse interactions with  various photon echo experiments.
Although the two-pulse photon echo itself cannot be used for quantum storage \cite{PhysRevA.79.053851}, the developed pulse area approach provides intensity independent universal tool for deeper studies of quantum memory (especially for intensive light pulses and cavity assisted storage), coherent spectroscopy and generation of  nonlinear waves in optically dense media.
It could also be used in both optical and microwave wavelength regions, for two- and three-level atomic ensembles with arbitrary transverse and longitudinal relaxation times, etc. 
Next important analytic step could be to generalize and extend the results acquired here for multi-pulse excitation using inverse scattering transform, as was done in \cite{Kaup1977} for McCall-Hahn area theorem. 

The reported study was funded by Russian Foundation for Basic Research, research project no.17-52-560009.
    
\bibliography{main}
\bibliographystyle{apsrev4-1}
    
    \newpage
    
\section*{Supplemental material}
    
 \subsection{Arbitrary echo pulse area}
Here we derive the general equation for an arbitrary individual echo pulse area. 
To do so we integrate the first of Eqs.~(1) over time around the time of echo emission $t_e$, from $t_0 = t_e-\tau/2$ to $t_1=t_e+\tau/2$, where $\tau$ is the delay between the pulses.
By assuming that $\tau \gg \delta t$, $\delta t$ being the pulse duration, we arrive to the equation for pulse area where we substitute the formal solution for $P$ from Eqs.~(1):
\begin{multline}
\partial_z \theta = i\frac{\mu}{2}\langle \int_{t_0}^{t_1} dt \Big[ P_0(\Delta) e^{-\gamma t_e-i \Delta (t-t_e)} 
\\
-i \int_{t_0}^{t} dt' \Omega (t') w(t',\Delta) e^{-(i \Delta + \gamma) (t-t')} \Big] \rangle,
\label{eq:area_1}
\end{multline}
where we introduced $P_0(\Delta)e^{-\gamma t_0} = P(t_0,\Delta)e^{-i \Delta \tau/2}$.

The key to finding the correct solution is proper handling of the integrals over $t$ in these two terms. 
One can show that $P_0(\Delta)$ and $w(t,\Delta)$ can be presented as a sum of several components $P_0 = P_0^{(0)}+P_0^{(1)}+...$ and $w = w^{(0)}+w^{(1)}+w^{(2)}+...$ with the total number of the components depending on the echo signal of interest (see Eqs. \eqref{eq:App_v_All} and \eqref{eq:App_w_All} and the following discussion). 
These components have a from $P_0^{(n)} \sim \exp[-i\Delta (t-t_e) -i n\Delta\tau + \varphi_n]$, $w^{(n)} \sim  \cos[n\Delta\tau+\varphi_n], \text{ where } n\in \mathbb{Z},$ the phase $\varphi_n$ is either $0$ or $\pi/2$.

For $n \neq 0$, $P_0^{(n)}$ and $w^{(n)}$ are rapidly oscillating functions of $\Delta$ near the echo pulse emission time $t_e$ since $\tau\gg \delta t$.
Averaging over $\Delta$ leads to that only the slowly varying terms $P^{(0)}_0$ and $w^{(0)}$ contribute to the echo pulse area in Eq.~\eqref{eq:area_1}.
After using $P_0(\Delta) = P_0^{(0)}$, we simply integrate the first term by taking into account: $\int_{t_0}^{t_1} dt e^{-i \Delta (t-t_e)} \rightarrow 2\pi \delta(\Delta)$ (this limit is valid assuming no temporal overlapping between the light pulses).
In the second term we switch the order of temporal integrals, similar to \cite{allen1975optical,Eberly:98}, and arrive to the integral:
\begin{align}
\begin{split}
    & \langle \int_{t_0}^{t_1} dt' \Omega (t') w(t',\Delta) \int_{t'}^{t_1} dt e^{-(i \Delta + \gamma) (t-t')} \rangle =
    \\
    & \langle \int_{t_0}^{t_1} dt \Omega (t)\tfrac{ w^{(0)}(t,\Delta)}{\gamma+i \Delta } \rangle = \pi G(0)\int_{t_0}^{t_1} dt\Omega (t)w^{(0)}(t,0),
\end{split}
\end{align} 
where we have also taken into account that $w^{(0)}(t',\Delta)$ and $G(\Delta)$ are even functions of $\Delta$. 
Thus Eq.~\eqref{eq:area_1} comes to: 
\begin{equation}
\partial_z \theta = \frac{\alpha}{2} \left[ 2  \tilde{v}_0
 +\int_{t_0}^{t_1} dt \Omega(t) \tilde{w}(t) \right],
\label{eq:area_3}
\end{equation}
where $\alpha = \mu\pi G(0)$ is the resonant absorption coefficient, $ \tilde{v}_0=iP_0^{(0)}(0) e^{-\frac{1}{2}\gamma\tau}$ is the resonant component of the phased  coherence, $\tilde{w}(t) = w^{(0)}(t,0)$ is the resonant component of the atomic inversion.
To find $\tilde{w}(t)$ and to integrate Eq.~\eqref{eq:area_3}, we write the Bloch equation set for the case $\Delta = 0$, ignoring relaxation during the pulses, since $\gamma \delta t\ll 1,$: 
\begin{align}
\begin{split}
    & \tilde{v}(t) = \tilde{v}_0 \cos \theta(t) + \tilde{w}_0 \sin \theta (t),
        \\
    & \tilde{w}(t)  = \tilde{w}_0 \cos \theta(t) - \tilde{v}_0 \sin \theta (t),
\end{split}
\label{eq:delta_sol}
\end{align}
where  $\theta(t) = \int_{t_0}^t \Omega(t) dt$, and $\tilde{v}(t)$ is a resonant part of the phased coherence,  $\tilde{w}_0 = \tilde{w}(t_0)$. 

Equation \eqref{eq:area_3} can now be integrated, and after reassigning $\tilde{v}_0 \rightarrow v_0,~\tilde{w}_0 \rightarrow w_0 $we obtain Eq. (5):
\begin{equation}
\partial_z \theta (z)= \frac{1}{2}\alpha [ 2 v_0(z) \cos^2\frac{\theta(z)}{2} + w_0(z)\sin\theta(z)],
\label{eq:area_general}
\end{equation}

 \subsection{Phasing components of polarization and inversion}
Here we show in detail the calculation of $v_0$ and $w_0$ for the secondary echo and give the expressions for the third and the fourth echoes.
We assume that the medium is excited by two incoming pulses having pulse areas $\theta_1, \theta_2$, that give rise to multiple photon echoes having pulse areas $\theta_{ei}$.

Under a multi-pulse excitation a two level system engages in two processes: it is either interacting with the electric field of the applied pulse, or it is left to its own devices and  experiences free oscillations decaying as $e^{-\gamma t}$. 
In the assumed timescales of these processes the influence of the pulse with area $\theta$ can be written as a rotation of the Bloch vector around $u$-axis:
\begin{equation}
T(\theta)\vec{r}=
\begin{pmatrix}
      1 & 0 & 0 \\
       0 & \cos \theta & \sin\theta \\
       0 & -\sin\theta & \cos\theta
    \end{pmatrix}
    \begin{pmatrix}
        u \\
        v \\
        w
      \end{pmatrix}.
\end{equation}
And free nutation is described with another rotation matrix, this time around $w$-axis:
        \begin{equation}
      U(t)\vec{r}=
    \begin{pmatrix}
       e^{-\gamma t} \cos \Delta t & - e^{-\gamma t} \sin  \Delta t & 0 \\
       e^{-\gamma t} \sin  \Delta t & e^{-\gamma t}  \cos  \Delta t & 0 \\
       0 & 0 & 1
    \end{pmatrix}
    \begin{pmatrix}
        u \\
        v \\
        w
      \end{pmatrix},
\end{equation}
    
    The secondary echo is emitted at the time $t=3\tau$, and to find the phasing parts of the coherence and inversion we write the Bloch vector 
    $\vec{r}(t)=U(t-2\tau)T(\theta_{e1})U(\tau)T(\theta_2)U(\tau)T(\theta_1) \vec{r}(0), t_0=5\tau/2$. 
    The calculation gives:
    \begin{align}
    \begin{split}
        v(t) = & - \Gamma_\tau c_1 c_2 s_{e1} c_t 
        \\
         & - \Gamma_\tau^2 [c_1 s_2 c_{e1} c_\tau c_t   + c_1 s_2 s_\tau s_t - s_1 s_2 s_{e1} c_t c_\tau] 
        \\
         +  \Gamma^3_\tau s_1 [ & c_\tau s_\tau s_t + c_2 c_\tau s_\tau s_t -  c_2 c_{e1} c_\tau^2 c_t +  c_{e1} s_\tau^2 s_t],
    \end{split}
    \label{eq:App_v_All}
        \\
        \begin{split}
        w(t) = & - c_1 c_2 c_{e1} + \Gamma_\tau [s_1 s_2 c_{e1} c_\tau+c_1 s_2 s_{e1} c_\tau ] 
        \\
        & -\Gamma_\tau^2 [s_1 c_2 s_{e1} c^2_\tau - s_1 s_{e1} s_\tau^2]
        \end{split}
        \label{eq:App_w_All}
    \end{align}
    here we use a short notation for trigonometric functions: $s_i = \sin \theta_i, c_i = \cos \theta_i, i=1,2,e1,$ $s_\tau = \sin \Delta\tau,$ $c_\tau = \cos \Delta\tau, s_t = \sin \Delta(t-\tau), c_t = \cos \Delta(t-\tau)$.
    
    This includes the phasing components, responsible for the echo generation and that are proportional to  $ \cos [\Delta (t-2\tau)]$ and non phasing components. 
    For example the first term of $v(t)$ contains only $c_t = \cos \Delta (t-\tau)$ and is non phasing, while the second term contains $c_\tau c_t = \cos \Delta\tau \cos \Delta(t-\tau) = \frac{1}{2}[\cos\Delta(t-2\tau)+\cos\Delta t] = \frac{1}{2}\cos\Delta(t-t_e)+\frac{1}{2}\cos(\Delta (t-t_e)+2\Delta\tau)$, so we get a phasing term $-\frac{1}{2}\Gamma_\tau^2  c_1 s_2 c_{e1} \cos\Delta(t-2\tau)$ that contributes to $P^0$.
    For $w(t)$ it is similar, except we are now interested in the time independent terms, like the first term in Eq.~\eqref{eq:App_w_All}. The terms with $c_\tau^2$ or $s_\tau^2$ also contribute since $c_\tau^2 (s_\tau^2) = \frac{1}{2}(1 \pm \cos 2\Delta\tau),$ where the second term will vanish after averaging over $\Delta$.
    
\begin{figure}
        \centering
        \includegraphics[width = \linewidth]{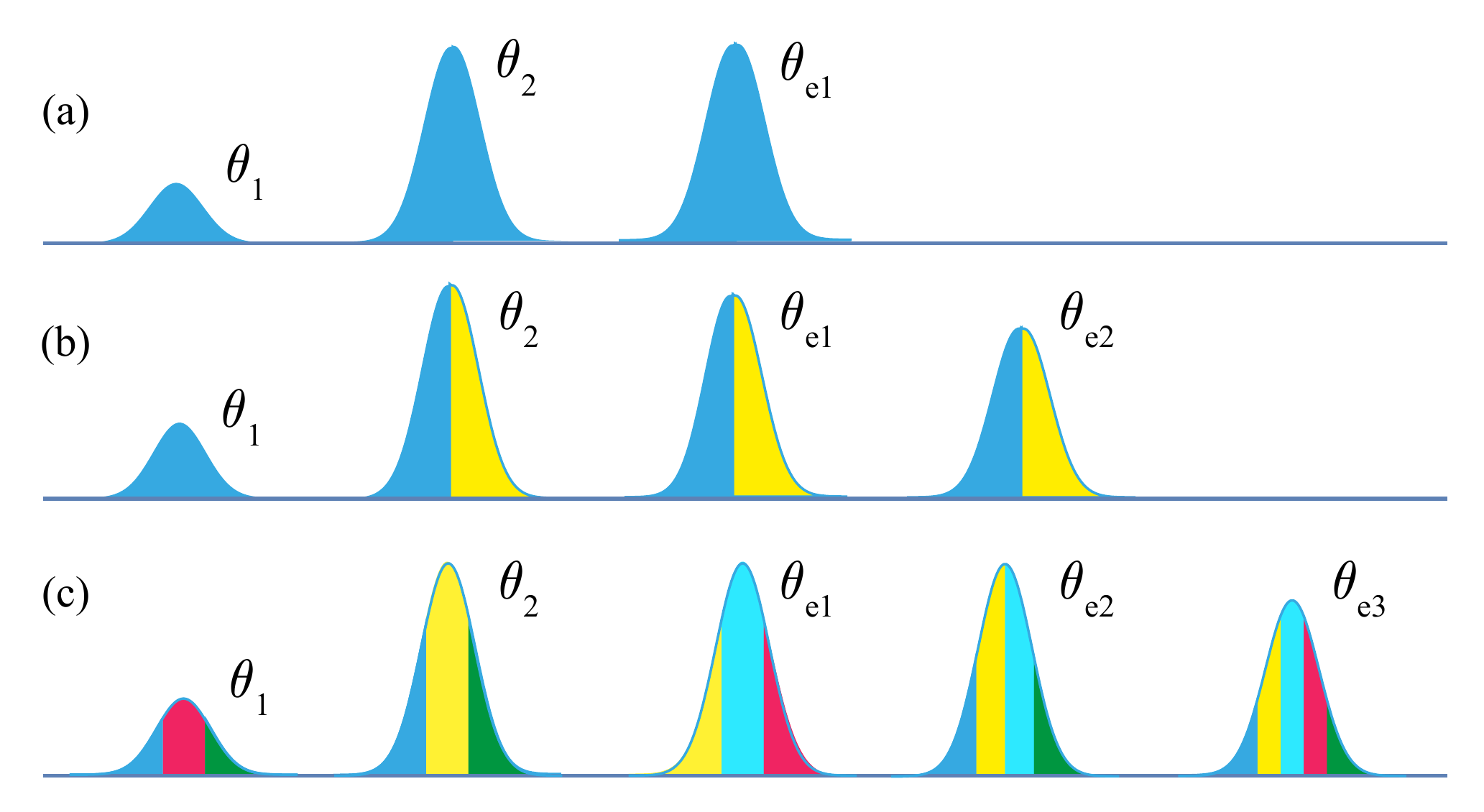}
        \caption{The diagram representing different contributions to the echoes. (a) The two-pulse (primary) echo. (b) The secondary echo has two contributions in source terms: the stimulated echo contribution is marked as blue and the two pulse echo contribution from $\theta_2, \theta_{e1}$ is marked as yellow. (c) Contributions to the echo emitted at $t=4\tau$; the five colors, blue, yellow, cyan, red and green respectively correspond to the five terms in $\tilde{v}_0(7\tau/2),$ Eq.~\eqref{eq:App_3rd_Echo_V}.}
        \label{fig:App_Pulse_Sequences}
    \end{figure}

    We now leave only the terms that contribute to the echo:
    \begin{align*}
        v(t) = & \frac{1}{2}\Gamma_\tau^2 [-(c_{e1} +1)c_1 s_2 + s_1 s_2 s_{e1}] \cos \Delta (t-2\tau), 
        \\
        w(t) = & - c_1 c_2 c_{e1} + \frac{1}{2}\Gamma_\tau^2 [1-c_2 ] s_1 s_{e1},
    \end{align*}
    and we get for $\tilde{v}_0(3/2\tau)$ and $\tilde{w}_0(3/2\tau)$:
    \begin{align}
        \begin{split}
            v_0(3/2\tau,z) = & -\Gamma^2_\tau \cos\theta_1 \sin\theta_2 \cos^2\tfrac{\theta_{e1}}{2} 
            \\
            & + \tfrac{1}{2} \Gamma^2_\tau \sin\theta_1 \sin\theta_2 \sin\theta_{e1},
        \end{split}
        \label{eq:App_Sec_V}
        \\
        \begin{split}
            w_0(3/2\tau,z) = & -\cos\theta_1 \cos\theta_2 \cos\theta_{eq} 
            \\
            & + \Gamma^2_\tau \sin\theta_1 \cos^2\tfrac{\theta_2}{2} \sin\theta_{e1}.
        \end{split}
        \label{eq:App_Sec_W}
    \end{align}
    The first terms in Eqs.~\eqref{eq:App_Sec_V} and \eqref{eq:App_Sec_W} are very similar to those of primary echo pulse and correspond to the two-pulse echo generation by the $\theta_2(z), \theta_{e1}(z)$. 
    This contribution to the secondary echo is presented as color yellow in Fig.~\ref{fig:App_Pulse_Sequences}. 
    The second terms in Eqs.~\eqref{eq:App_Sec_V},\eqref{eq:App_Sec_W} correspond to the stimulated echo generation and are presented by the color blue in Fig.~\ref{fig:App_Pulse_Sequences}.
  
    In the same fashion we can write the phasing coherence and inversion after four pulses, two incoming pulses and two echo pulses: 
    \begin{align}
        \begin{split}
            v_0 (7\tau/2,z) = & \tfrac{1}{2} \Gamma_\tau^2 \times
            \\
            & \big[\sin\theta_1\sin\theta_2\cos\theta_{e1}\sin\theta_{e2} \\
            & +\cos\theta_1\sin\theta_2\sin\theta_{e1}\sin\theta_{e2} \\
            & +2\cos\theta_1\cos\theta_2\sin\theta_{e1}\sin^2\tfrac{\theta_{e2}}{2}\big]
        \\
            & + \Gamma_\tau^4    \big[\sin\theta_1\cos^2\tfrac{\theta_2}{2}\sin^2\tfrac{\theta_{e1}}{2}\cos^2\tfrac{\theta_{e2}}{2} \\
            & -\sin\theta_1 \sin^2\tfrac{\theta_2}{2}\cos^2\tfrac{\theta_{e1}}{2}\sin^2\tfrac{\theta_{e2}}{2} \big],
       \end{split}
       \label{eq:App_3rd_Echo_V}
       \\
       \begin{split}
               w_0(7\tau/2,z) = & -\cos\theta_1\cos\theta_2\cos\theta_{e1}\cos\theta_{e2} \\
            & - \Gamma_\tau^2 \times [ \sin\theta_1\sin^2\frac{\theta_2}{2}\sin\theta_{e1}\cos\theta_{e2} \\
            & + \cos\theta_1\sin\theta_2\sin^2\frac{\theta_{e1}}{2}\sin\theta_{e2}  \\
            & + \frac{1}{2}\sin\theta_1\sin\theta_2\sin\theta_{e1}\sin\theta_{e2} ] = \\
            & -v_4 \sin\theta_{e2} + w_4 \cos\theta_{e2}.
        \end{split}
        \label{eq:App_3rd_Echo_W}
    \end{align}

    The first term $\sim \sin\theta_1\sin\theta_2\cos\theta_{e1}\sin\theta_{e2}$ in $ v_0 (7\tau/2,z)$ is the stimulated echo from the two incoming and the second echo pulses naturally proportional to $\Gamma_\tau^2$. 
    The second term $\sim \cos\theta_1\sin\theta_2\sin\theta_{e1}\sin\theta_{e2}$ is another stimulated echo generated by the second incoming and the first two echo pulses. 
    The third term $\sim\cos\theta_1\cos\theta_2\sin\theta_{e1}\sin^2\tfrac{\theta_{e2}}{2}$ represents the contribution of the two-pulse echo from the two echo pulses.
    The next term $\sim \sin\theta_1\cos^2\tfrac{\theta_2}{2}\sin^2\tfrac{\theta_{e1}}{2}\cos^2\tfrac{\theta_{e2}}{2}$ is the two-pulse echo generated by the first incoming and primary echo pulses.
    The last term, $\sim \sin\theta_1 \sin^2\tfrac{\theta_2}{2}\cos^2\tfrac{\theta_{e1}}{2}\sin^2\tfrac{\theta_{e2}}{2}$ is the revived primary echo, generated by the first two incoming pulses and recovered  by the second echo pulse (first echo pulse just suppresses its amplitude by the factor $\cos^2\tfrac{\theta_{e1}}{2}$).

\end{document}